\documentstyle[11pt,paspconf,epsf]{article}

\begin{document}

\title{A New Object-Weighted Measure of the Small-Scale Velocity
  Dispersion}

\author{Jonathan E.~Baker and Marc Davis}

\affil{Astronomy Department, University of California, Berkeley, CA
  94720}

\begin{abstract}
  We describe a new statistic for measuring the small-scale velocity
  dispersion of galaxies directly from redshift surveys.  This
  statistic is based on the object-weighted statistic proposed by
  Davis, Miller, \& White (1997).  Compared with the traditional pair-weighted
  velocity dispersion, our statistic is less sensitive to the presence
  or absence of rare, rich clusters of galaxies.  This measure of the
  thermal energy of the galaxy distribution is ideally suited for use
  with a filtered version of the cosmic energy equation.  We discuss
  the application of the statistic to the Las Campanas Redshift
  Survey.  The low observed dispersion strongly favors cosmological
  models with low matter density, $\Omega_m\sim 0.2$.
\end{abstract}

\section{Introduction}

Redshift surveys provide an accurate measure of the Hubble recession
velocity $H_0 r$ plus radial peculiar velocity (relative to the
velocity $\mathbf{v}_0$ of the observer):
\begin{equation}
  \label{eq:cz}
  cz = H_0 r + (\mathbf{v}_{\mathrm{pec}} - \mathbf{v}_0) \cdot
  \hat{\mathbf{r}}
\end{equation}
for large samples of galaxies.  In the gravitational instability
paradigm for structure formation, peculiar velocities grow in response
to the total (not just visible) amount of clustered mass.
Measurements of the magnitude of these peculiar velocities are an
important cosmological probe, with the potential to discriminate
cosmological models, to constrain the bias of the galaxy distribution,
and to constrain models for structure formation.

On relatively large (few Mpc and up) scales, peculiar velocities
define a smooth flow field, the divergence of which is simply given by
the galaxy over-density times the parameter $\beta\approx
\Omega_m^{0.6}/b$, where $b=\delta_g/\delta_m$ is the galaxy bias.  A
number of methods have been devised for combining redshift and
peculiar velocity surveys to obtain $\beta$ from these large-scale
flows (Strauss, this volume; Willick, this volume).  These methods
have not fully converged to a consistent solution; while POTENT
prefers $\beta\sim 1$ (e.g., {Sigad} {et~al.} 1998), other methods tend to
give low values $\beta\approx 0.5$ (e.g., {Willick} \& {Strauss} 1998).

Turning to small scales of order 1 Mpc, we find that peculiar
velocities are essentially thermal or incoherent.  The kinetic energy
contained in these motions can be used with a filtered version of the
cosmic energy equation to yield an estimate of $\Omega_m/b^2$,
which is approximately equal to $\beta^2$.

It has been known for some time that the thermal energy of the galaxy
distribution is quite low, or equivalently the cosmic Mach number is
rather high, relative to our theoretical expectations.  A simple way
to see this is to take an $N$-body simulation and plot what it would
look like in redshift space.  In a box of length 5000 km s$^{-1}$, one
sees very prominent, long ``fingers of god'', which tend to wash out
the collapsed filamentary structures typical of real redshift surveys.
The flow in the vicinity of the Local Group has also been measured to
be remarkably cold, with a dispersion of only 60 km s$^{-1}$, and no
blue-shifted galaxies are seen outside the Local Group
(Schlegel, Davis, \& Summers 1994).  Governato {et~al.} (1997) showed that this is
extremely hard to reproduce even in low-$\Omega_m$ $N$-body
simulations.

The pair velocity dispersion ($\sigma_{12}$) is the traditional
measure of small-scale velocities.  Redshift surveys provide a set
$\{r_p, \pi\}$ of projected and radial separations in redshift space;
$\sigma_{12}$ can be estimated directly from the redshift-space
correlation function $\xi_z(r_p, \pi)$, which is a convolution of the
real-space correlation function $\xi(r)$ with the pair velocity
distribution function $f(v)$ (the rms of which is $\sigma_{12}$).  An
exponential $f(v)$ has been found to fit well and is also expected
from theoretical considerations.

The $\sigma_{12}$ statistic was first applied to the CfA redshift
survey by Davis \& Peebles (1983), who measured 340 km s$^{-1}$ at $1h^{-1}$
Mpc scales.  The fact that this dispersion was much lower than that of
the $\Omega_m=1$ simulations of Davis {et~al.} (1985) was the original
motivation for bias in the galaxy distribution.
{Gelb} \& {Bertschinger} (1994) later showed that no normalization of the
standard Cold Dark Matter (SCDM) model could simultaneously match both
the observed correlation amplitude and velocity dispersion.  However,
many authors pointed out that the pair statistic was not robust (e.g.,
Mo, Jing, \& Borner 1993; Zurek {et~al.} 1994; Somerville, Primack, \&  Nolthenius 1997;
Guzzo {et~al.} 1997; Jing, Mo, \& Borner 1998).  In particular, its pair-wise
weighting makes $\sigma_{12}$ very sensitive to the presence or
absence of rare, rich clusters in the survey volume; the treatment of
a few objects can greatly affect the result.  The mean streaming
motions of galaxies were also shown to have a considerable effect on
the measured value of the dispersion.

In this work, we present the application of a new, object-weighted
statistic to the Las Campanas Redshift Survey
(LCRS; Shectman {et~al.} 1996).  The LCRS is the largest existing
redshift survey which is nearly fully sampled.  It contains 26,418
galaxies with a median redshift of approximately 30,000 km s$^{-1}$.  The
survey consists of six $1\fdg5\times 80\deg$ slices (three at
northern declinations and three in the south), and contains about
30 clusters.

Even with such a large and well-sampled survey as the LCRS, the
measurement of the pair dispersion $\sigma_{12}$ has been fraught with
controversy.  Fourier techniques (Landy, Szalay, \& Broadhurst 1998) gave a low
dispersion consistent with the old CfA value, but
{Jing} \& {Borner} (1998) showed that accounting for mean streaming
motions increased the value to $570\pm 80$ km s$^{-1}$.  Jing {et~al.} (1998)
found that simulations with $\Omega_m h \sim 0.2$ could reproduce the
observed pair dispersion, but they required a somewhat mysterious
anti-bias in the galaxy distribution (i.e., higher mass-to-light
ratios in dense regions).

Given the difficulty of reliably estimating the pair dispersion, it is
clearly of interest to develop alternative statistics.  Our statistic,
called $\sigma_1$, is a modified version of the object-weighted
statistic developed by Davis, Miller, \& White (1997, hereafter DMW).  We note that
other alternatives to $\sigma_{12}$ have been suggested.
Nolthenius \& White (1987) proposed the mean dispersion of groups of
galaxies; our statistic has the advantage that it does not require
assigning galaxies to groups and averaging over the internal motions.
The pair dispersion can also be measured as a function of local
density (Kepner, Summers, \& Strauss 1997; {Strauss}, {Ostriker}, \&  {Cen} 1998), but unlike $\sigma_1$,
this must be computed in volume-limited samples.  This work discusses
the application of $\sigma_1$ to the LCRS; a more complete description
of our results may be found elsewhere (Baker, Davis, \& Lin 1999).

\section{The $\sigma_1$ Statistic}

We start from the definition of the single-particle object-weighted
dispersion proposed by DMW.  These authors applied this statistic to
the UGC and \textsl{IRAS} 1.2-Jy redshift surveys, which are much
smaller than the LCRS survey.  For the UGC catalog, they measured
$\sigma_1 = 130\pm 15$ km s$^{-1}$, much colder than their $\Omega_m=1$
$N$-body simulation, even when the simulation velocities were
artificially cooled by a factor of two.

\subsection{Definition of $\sigma_1$}

Around each survey galaxy $i$ in redshift space, we place a cylinder
of radius $r_p$ and half-length $v_l$, with the axis of symmetry along
the redshift (radial) coordinate.  All the galaxies which fall within
the cylinder are considered to be neighbors of galaxy $i$.  We
typically take $r_p=1h^{-1}$ Mpc and $v_l=2500$ km s$^{-1}$.  We
construct a histogram of the number of neighbor galaxies $P_i(\Delta
v)$ in bins of velocity separation $\Delta v$.  We subtract the
background distribution $B_i(\Delta v)$ expected for an uncorrelated
galaxy distribution, and then average over the $N_g$ galaxies to
obtain the final distribution:
\begin{equation}
  \label{eq:ddv}
  D(\Delta v) = \frac{1}{N_g} 
  \sum_i w_i \left[ P_i(\Delta v) - B_i(\Delta v) \right],
\end{equation}
where the weight for galaxy $i$ is denoted by $w_i$.

In the original DMW formulation, the weight for each galaxy was simply 
given by its total number of neighbors in excess of background:
\begin{equation}
  \label{eq:whi}
  w_i^{-1} = \sum_{\Delta v} \left[ P_i(\Delta v) - B_i(\Delta v) \right].
\end{equation}
It is this factor which gives the statistic its essential object
weighting, in contrast to the traditional pair-weighted dispersion.
However, galaxies with fewer neighbors than the background had to be
deleted from consideration, which biased the statistic towards higher
density (and thus hotter) objects.

\begin{figure}[tb]
  \plotone{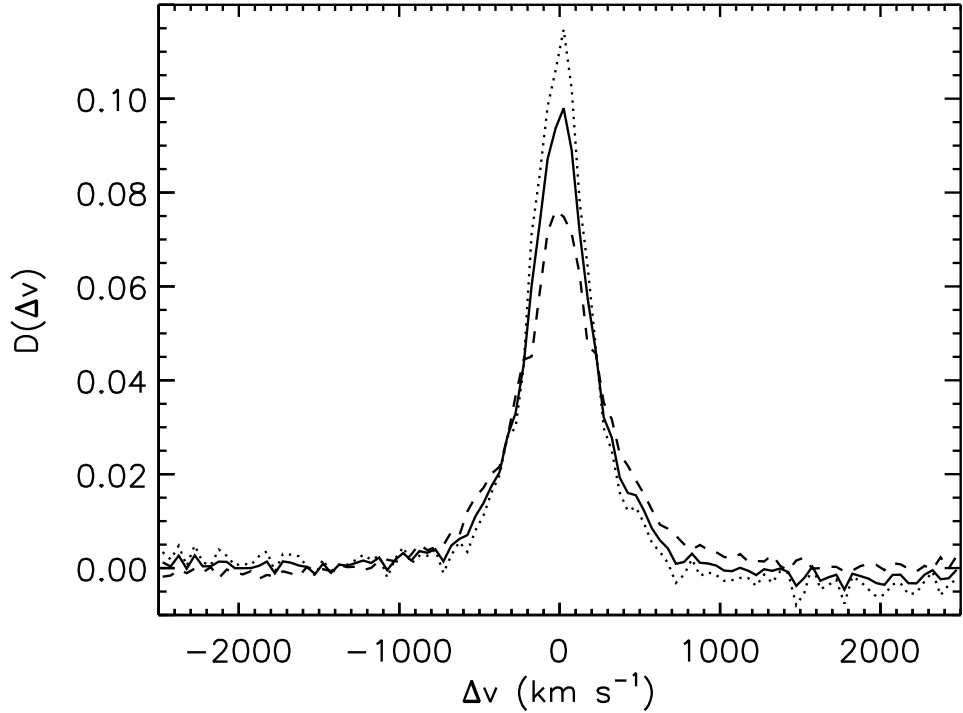}
  \caption{Object-weighted velocity distributions for the LCRS.  We
    show distributions for galaxies in low-density (dashed) and
    high-density (dotted) regions, and the combined distribution
    (solid).  The widths are $\sigma_1=$ 99, 207, and 126 km s$^{-1}$,
    respectively.}
  \label{fig:hilo}
\end{figure}

We avoid the bias inherent in the original statistic by assigning
$w_i=1$ for galaxies with less than one excess neighbor.  In order to
combine the high- and low-density galaxies sensibly, it is necessary
to tabulate their distributions separately.  We subtract off the tails
of the distributions (within 500 km s$^{-1}$ of $v_l$) and then normalize
each distribution so that the sum $\sum_{\Delta v} D(\Delta v)$ is
proportional to the number of galaxies included.  This allows the
high- and low-density distributions to be combined so that they are
weighted according to the number of objects assigned to them (see
Figure~\ref{fig:hilo}).

We measure the width $\sigma_1$ of the resulting distribution by
performing a $\chi^2$ fit to a model
\begin{equation}
  \label{eq:model}
  M(\Delta v) = \overline{\xi_R} \ast f \ast E.
\end{equation}
Here $\overline{\xi_R}$ is the two-point correlation function $\xi(r)$
averaged in the cylindrical bins, $f\propto e^{-|v|/\sigma_1}$ is the
velocity distribution function, and $E$ is the error distribution of
the redshift measurements (for the LCRS, a Gaussian of rms 67 km s$^{-1}$).
We find that the exponential form for $f$ fits much better than a
Gaussian for both the data and $N$-body simulations.  We have defined
$\sigma_1$ so that it is a measure of the one-dimensional dispersion
of the motion of individual galaxies, with bulk flows filtered out on
scales larger than $1h^{-1}$ Mpc.  We note that the rms of the distribution
$f$ is $\sigma_1 \sqrt{2}$, which DMW distinguished from $\sigma_1$ by
calling it $\sigma_I$.

\subsection{Cosmic Energy Equation}

The cosmic energy, or Layzer-Irvine, equation is a differential
relation between the kinetic and potential energies of the mass
fluctuations in an expanding universe.  It has been shown that for
self-similar cosmological clustering, the equation reduces to an
algebraic expression: $\langle v_{\mathrm{pec}}^2 \rangle \approx g
\Omega_m H_0^2 J_2$, where $J_2\equiv\int r\xi(r)\, dr$ measures the
potential energy and $g\approx 0.3$.

The difficulty in applying this equation to measure the mass density
arises because the kinetic energy term and $J_2$ are very poorly
constrained on large scales.  We therefore consider a filtered version
of the equation, including contributions only from small scales.  The
simulations of DMW showed that $\sigma_1$ is a good measure of the
small-scale kinetic energy and can be used to estimate $\Omega_m/b^2$
(the factor of $b^2$ arises because we can only measure $J_2$ for the
galaxies rather than the underlying mass).

\subsection{$N$-body Simulations} 

We have completed several cluster-normalized $N$-body simulations for
comparison with the LCRS.  The cosmological parameters of our models
are given by Baker {et~al.} (1999).  They include a standard $\Omega_m=1$
(SCDM) model, a tilted variant with $n=0.8$ (TCDM), a flat model with
a cosmological constant $\Omega_\Lambda=0.7$ (LCDM), and an open model 
with $\Omega_m=0.3$ (OCDM).

The models were evolved using a P3M code (Brieu, Summers, \& Ostriker 1995) and a
special-purpose GRAPE-3AF board (Okumura {et~al.} 1993), which is
hardwired to compute the Plummer force law very quickly.  The board is
attached to a Sun SPARC workstation which computes the long-range
contributions to the forces.  We are able to complete one cosmological
run with $64^3$ particles on a $128^3$ mesh in approximately one
CPU-day.  Our box size is $L=50h^{-1}$ Mpc to match the length of the
redshift-space cylinders used in the $\sigma_1$ analysis.

We apply the same statistical procedure for computing $\sigma_1$ for
the LCRS to the particles in the simulations.  Of course, galaxies may
in general have a different velocity dispersion from the underlying
mass, and the problem of identifying ``galaxies'' in the simulations
is therefore an important one.

We first apply the standard friends-of-friends algorithm for defining
halos, with a linking length of 0.2 mesh cells.  With our relatively
poor mass resolution, this procedure leads to a serious and well-known
over-merging problem, yielding halo correlation functions which are
much too low on small scales.  To remedy this situation, we subdivide
large ($N>N_s$) halos by drawing individual particles from them at
random, with a probability $\propto N^{-\alpha}$.  For $\alpha>0$,
this yields a mass-to-light ratio which increases at small scales;
this is required because $\alpha=0$ leads to correlation functions
which are too steep in many models.  

We choose the parameters $\alpha$ and $N_s$ to yield ``galaxies''
which match the correlation function and number density of the LCRS
(about 2500 per simulation volume).  We find that $\alpha\sim 0.25$ is
typically the best choice, and $N_s\sim 80$, corresponding to a mass
of $10^{13}\Omega_m h^{-1}$ M$_\odot$.  Although by selecting
individual particles as galaxies we are including the internal
dispersions of galaxies, these velocities are small compared to the
dispersions of the large clusters which are split.  We have also
measured $\sigma_1$ for galaxies drawn from a large Virgo simulation
({Benson} {et~al.} 1999) using semi-analytic techniques, and we find results
consistent with our simulations.

\section{Results}

\begin{figure}[tb]
  \plotone{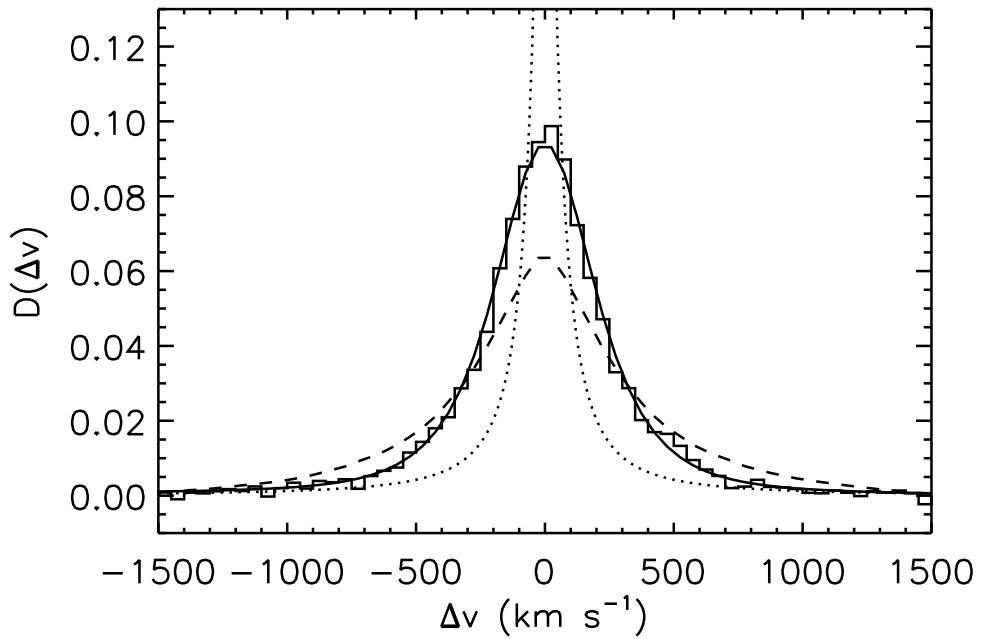}
  \caption{Velocity distributions for the LCRS data (histogram) and
    best-fitting model (solid curve) with $\sigma_1=126$ km s$^{-1}$.  Also
    shown are a model with no broadening ($\sigma_1=0$; dotted) and
    the distribution for the mass in our SCDM simulation
    ($\sigma_1=310$ km s$^{-1}$; dashed).}
  \label{fig:dv}
\end{figure}

Based on the mean of the six LCRS slices, we measure $\sigma_1 =
126\pm 10$ km s$^{-1}$, where the quoted error is the standard deviation of
the mean for the slices.  The statistic is quite robust, with
similar $D(\Delta v)$ distributions for each of the slices.  The fit
is quite good, with $\chi^2_\nu = 117/96 = 1.22$, where we have
estimated the errors from the standard deviation of the six slices.
The distribution is plotted in Figure~\ref{fig:dv}.

The exponential model also provides an excellent fit to the $D(\Delta
v)$ distributions in the simulations.  We find that the mass in the
SCDM and TCDM models is much too hot, with dispersions of over 300
km s$^{-1}$.  The LCDM and OCDM models yield lower dispersions $\sim 200$
km s$^{-1}$; the OCDM model is slightly hotter than the LCDM.  The halos are
somewhat cooler, with velocity biases in the range 0.7--0.9.  The LCDM 
halos are the best match to the LCRS dispersion, with $\sigma_1=143$
km s$^{-1}$.

Combining the LCRS and $N$-body halo results, we can solve the cosmic
energy equation for $\Omega_m$ (the bias factor drops out because we
have chosen halos which match the LCRS correlation function).  We
obtain similar results from each of the four cosmological models, with
$\Omega_m$ in the range 0.15--0.25.  If we combine the LCRS and
$N$-body mass, we obtain $\Omega_m^{0.5}/b$, approximately equal to
$\beta$.  The result is 0.3--0.4 for the two high-density models, and
0.4--0.6 for the two low-density models.

\begin{figure}[tb]
  \plotone{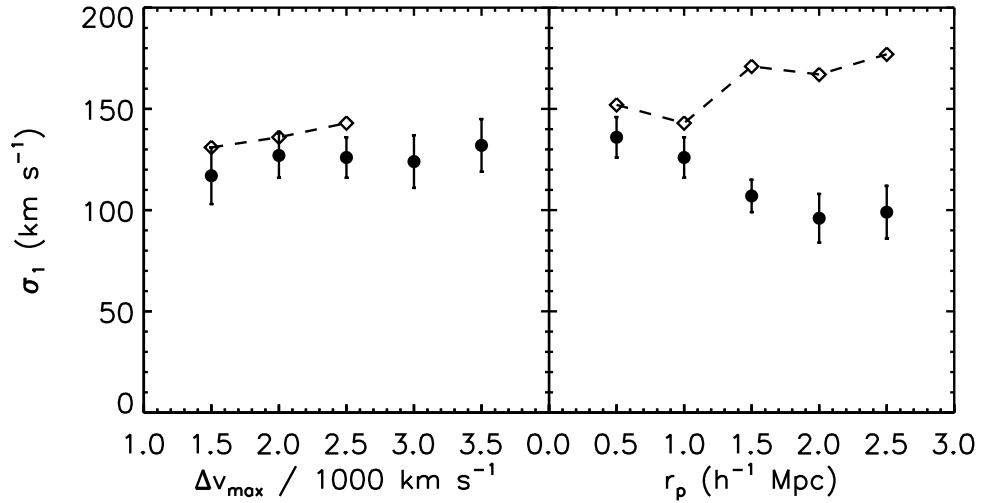}
  \caption{Dependence of $\sigma_1$ on the cylinder length and
    radius.  Filled points with error bars are for the LCRS, open
    points and dashed line are for galaxies in our LCDM simulation.}
  \label{fig:scale}
\end{figure}

As we increase the radius $r_p$ of the cylinders used to measure
$\sigma_1$, we find an interesting discrepancy between the models and
the LCRS data (Figure~\ref{fig:scale}).  
The LCRS $\sigma_1$ decreases modestly as the scale is increased, but
all of the models show an increase with scale.  It is at present
unclear whether this discrepancy is a numerical artifact of the
simulations or a real physical effect.

\section{Conclusions}

We have applied a new object-weighted, unbiased measure of the
small-scale velocity dispersion to the Las Campanas Redshift Survey.
We derive a single-particle dispersion $\sigma_1=126\pm 10$ km
s$^{-1}$.  Our statistic has considerable advantages over the
traditional pair dispersion $\sigma_{12}$; namely, it is less
sensitive to rare, rich clusters of galaxies.  Our statistic should
play an important role in analyses of future redshift surveys.

When compared with the LCRS data, cluster-normalized $\Omega_m=1$
$N$-body simulations are far too hot on small scales.  We find strong
evidence for a low density $\Omega_m\sim 0.2$, and we derive
consistent values of the density parameter from a number of different
models; these results are described more fully elsewhere
(Baker {et~al.} 1999).  In addition to constraining the mass density, the
$\sigma_1$ statistic applied to upcoming surveys should provide
important constraints on the galaxy bias and evolution of structure.

\acknowledgments

We would like to thank S.~Corteau for organizing a tremendously
successful and enjoyable conference in Victoria, with an abundance of
Nanaimo bars which were consumed rapidly by at least one of the
authors.  J.~E.~B. acknowledges the support of an NSF graduate
fellowship.  This work was also supported in part by NSF grant
AST95-28340.


\end{document}